\documentclass[smallextended]{svjour3}
\smartqed
\usepackage[pdftex]{graphicx}
\usepackage{mathrsfs}
\usepackage[colorlinks, breaklinks, urlcolor={blue}, linkcolor={red}, citecolor={blue}]{hyperref}
\usepackage[T1]{fontenc}
\usepackage{array}
\usepackage{amsmath}
\usepackage{type1cm}
\usepackage{lettrine}
\usepackage[english]{babel}
\usepackage{lmodern}
\usepackage{microtype}
\usepackage{booktabs}
\usepackage[boxed, vlined]{algorithm2e}
\usepackage{caption}
 \usepackage{braket}
\newsavebox{\definitionbox}
{\end{minipage}\end{lrbox}%
\begin{center}{\colorbox{checkcolor}{\usebox{\definitionbox}}}%
\end{center}}
\title{Multi-path Security in Quantum Networks with Trusted Nodes}
\author{Muhammad Kumail Haider$^\dag$\and Aeysha Khalique $^{\dag\ddag\mathparagraph}$}
\institute{$\dag$ \at School of Natural Sciences, National University of Sciences and Technology,
	H-12 Islamabad, Pakistan
    \and $\ddag$ \at Department of Physics, Syed Babar Ali School of Science and Engineering, Lahore University of Management Sciences, Lahore, 54792, Pakistan
    \and $\mathparagraph$ \at National~Centre~for~Physics~(NCP),~Shahdra~Valley~Road,~Islamabad~44000,~Pakistan.} 
\date{Received: date / Accepted: date}
\date{\today}
\begin{document}

\maketitle

\begin{abstract}
Quantum communication networks promise information-theoretic security but this security collapses if intermediate relay nodes are compromised, a risk that grows sharply when compromised nodes cannot be identified. We address this twofold: first, we show that multi-path routing provides quantifiable protection against unidentified compromised nodes. Second, we provide a comprehensive statistical and graph-theoretic analysis of network and node security under compromise, covering full and partial node compromise, optimal placement of additional edges via an iterative method, and comparative hacking-probability analysis across network topologies. Together, these results identify the topology and multi-path routing strategy that most improves security for a given network. We apply this framework to quantum key distribution across multiple network topologies, demonstrating a scalable route to secure, reliable quantum communication under realistic, unidentified adversarial threats, without relying on quantum memory.
\end{abstract}

\section{Introduction}
Quantum networking has become a key technology in pursuing many communication tasks like quantum key distribution~\cite{bennett2014,ekert1991quantum}, quantum teleportation~\cite{bennett1993teleporting} and distributed quantum computing~\cite{caleffi2018quantum}. The security of the quantum network is based on laws of physics rather than computational complexity. 

Quantum network has a distance limitation because the quantum signals (photons) lose coherence and get absorbed or scattered over long distances, limiting achievable distances~\cite{gisin2002quantum}. In entanglement based networks, quantum repeaters are deployed to reduce distance limitations~\cite{azuma2023quantum}. These repeaters use entanglement swapping and quantum error correction to maintain the integrity of quantum states over long distances~\cite{briegel1998quantum}. 
 Multi-path approach has been used with quantum repeaters to enhance the robustness and efficiency of long-distance quantum networking by reducing the likelihood of failure due to photon loss or noise on a single path~\cite{mondal2024multipath,leone2021qunet,daud2024scaling}. 
 
Entanglement-based schemes are difficult to implement because of the fragility and the degradation of fidelity of these states over long distances. Trusted relays provide a viable solution for a secure network over long distances, by utilizing prepare and measure schemes where there is no need to preserve entanglement~\cite{bechmann505089quantum}. Such nodes form an intermediate communication stage that extends quantum keys through networks while allowing communication systems to expand distances. Security relies heavily on these nodes being trusted.
In these networks, the security of the network can be completely undermined if an adversary compromises a node within the network~\cite{elliott2002building}. multi-path routing has emerged as the primary defense against this risk, and this approach is already being deployed in real systems~\cite{Solomons2022PRXQuantum}. Existing multi-path analyses have advanced routing efficiency~\cite{Kiktenko2024Entropy} and closed-form security bounds under various attack models~\cite{StepniakJISA2023}.

In this paper, we develop a statistical model for multi-path network security and use it to design algorithms that route securely around unidentified compromised nodes. These algorithms efficiently find the most secure independent path-set between two nodes, and guarantee the security given up to $T$ compromised nodes at unknown locations. We extend this analysis to partial compromise, where every node carries some nonzero probability of being compromised, and systematically compare security and resilience across seven realistic topologies: line, star, cycle, grid, tree, wheel, and connected-tree. Building on this, we propose an iterative method for identifying the optimal edge to add to a given topology under a limited resource budget, offering practical guidance for topology selection and network design. Finally, we demonstrate multi-path-routed QKD across these topologies using our statistical model, showing how key rate varies with node compromise and validating the framework end-to-end without requiring quantum memory. Together, these results offer a way to quantify, compare, and actively design for network security using multi-path routing under unidentified node compromise, contributing a scalable, deployment-ready step toward robust quantum communication networks.

The remainder of this paper is organized as follows: Sec.~\ref{Sec:qkd} provides a detailed overview of quantum communication schemes based on trusted nodes and the foundation of multi-path networking. Sec.~\ref{Sec:Multipath_sec} describes the methodology used in our study, including the statistical and graph theoretic modelling of a network, and the analysis of node compromise scenarios. We present the results of our analysis of the proposed multi-path framework for different network topologies in Sec.~\ref{sec: Analysis}. Sec.~\ref{Sec:Key_rate} presents the application of the network to quantum key distribution. Finally, we conclude in  Sec.~\ref{Sec: Conclusion}.
\section{Background}\label{Sec:qkd}
Long-distance quantum communication can be achieved by entanglement swapping and quantum-memory based repeater setup~\cite{abruzzo2014long}. While entanglement is fragile and quantum memories are experimentally challenging~\cite{ma2020optical}, an alternative is the classical repeater setup based on trusted nodes. We introduce trusted nodes setup in Sec.~\ref{sec:reptrusted} in comparison with quantum repeaters. In Sec.~\ref{sec:TrustedNodesSchemes}, we outline some communication schemes in a trusted node network. We introduce the foundation of multi-path networking in Sec.~\ref{sec:KDF}.

\subsection{Trusted Nodes vs Quantum Repeaters}\label{sec:reptrusted}
\begin{figure}[tbp]
    \includegraphics[width=\columnwidth]{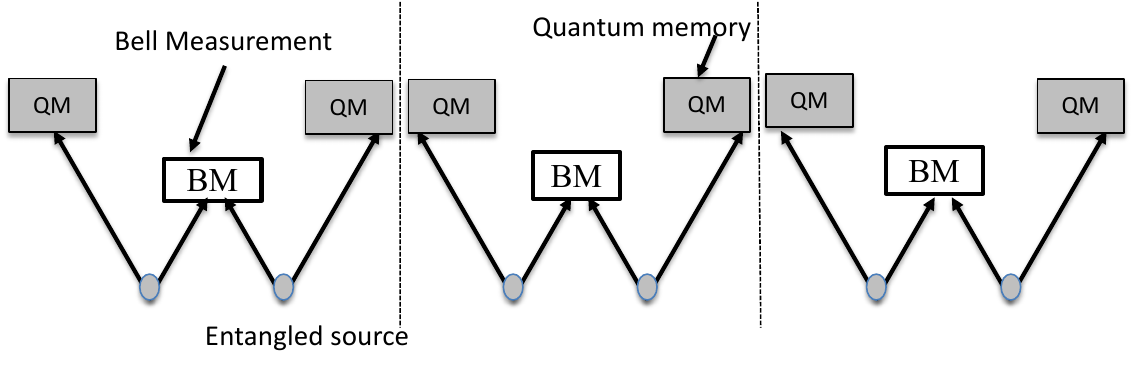}
    \caption{In a quantum repeater network long-range end-to-end entanglement is formed using elementary entanglement swapping and purification operations. If Alice and Bob are provided with a resource of entangled states, $\ket{\Psi}_{AB}$, this enables the information-theoretically secure sharing of random bit-strings between them, which is robust against man-in-the-middle attacks.}
    \label{fig:repeater_network}
\end{figure}

Quantum repeater setup is based on entanglement swapping and quantum memories as shown in Fig.~\ref{fig:repeater_network}.  The path length $L$ between the two users is divided into N vertices. Each vertex contains a source of entangled pair. One qubit from each node is kept in quantum memory and the other is sent to the Bell measurement apparatus together with one qubit from the adjacent vertex. Once all Bell measurements are done, the qubits from the quantum memories are taken out and Bell measurements are performed between the ones from adjacent memories, thus extending the distance until the two end users are entangled.~\cite{briegel1998quantum}

Quantum repeaters are experimentally challenging to achieve. The most important ingredient is quantum memory,~\cite{wolters2017simple} which is needed to store the qubit while preserving its entanglement. Although extensive efforts are being made to achieve quantum memories using ensembles or single atoms or ions in special traps or solid-state crystals, yet there are many challenges in using them in a network.~\cite{kimble2008quantum}

Given the challenge of storing and preserving entanglement, an alternate way may be a classical repeater setup (Fig.~\ref{fig:trusted_nodes_network}) using trusted nodes. The nodes are connected point to point through quantum channel and the quantum state is measured and then resent at each trusted node.

\subsection{The Trusted Nodes Quantum Networks}\label{sec:TrustedNodesSchemes}
 
Trusted relay schemes are a foundational approach in quantum networks. The relay is built as the key is distributed across the network by hop-by-hop encryption schemes where keys are pre-shared key $QK_i$ and between two adjacent nodes $i$ and $i+1$ using standard QKD techniques. Some of the schemes are explained below.

\begin{figure}[tbp]
\centering
         \includegraphics[scale=0.3]{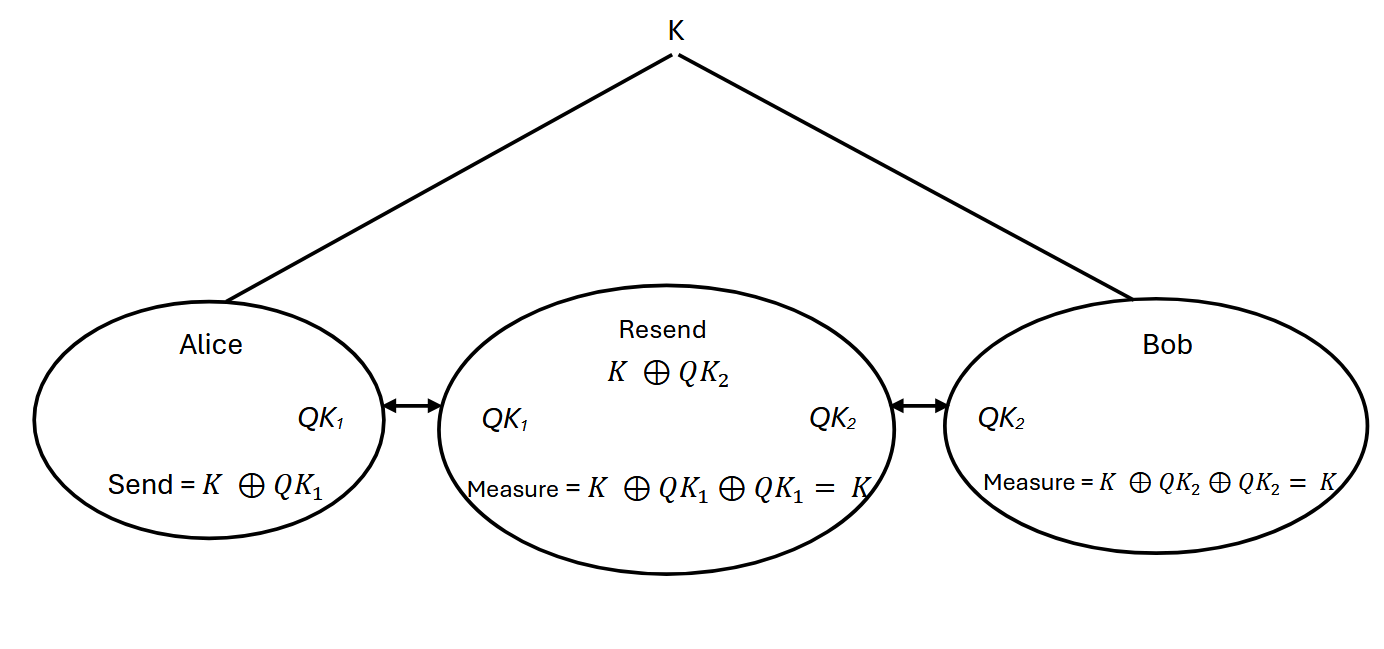}
  \caption
  {Trusted nodes setup with BBN QKD scheme. Key $K$ from Alice to Bob through an intermediate relay using shared keys $QK_1$ and $QK2$.
  }
     \label{fig:trusted_nodes_network}
 \end{figure}
The BBN trusted Relay scheme which was first proposed in 2002~\cite{elliott2002building}, has been in continuous operation since 2003 in the DARPA Quantum Network~\cite{elliott2004quantum}. The process is shown in Fig.~\ref{fig:trusted_nodes_network} and uses the pre-shared keys. The key $K$ to be shared is encrypted at the neighboring nodes using the shared key and then re-encrypted using the shared key with the next node. The process continues until the key $K$ reaches the end user. A somewhat similar re-encryption scheme~\cite{schartner2009overcome} offers more security than BBN by creating a new key $x_i=QK_i\oplus QK_{i+1}$ at each node. The cipher text $c_i$ received from the previous node is then re-encrypted using $c_i+x_i$ and send to the next node. In public XOR scheme~\cite{dong2019wide}, the $x_i$ are generated as in re-encryption scheme, and are announced at public channel. The end user takes XOR of all $x_i$ and his shared key $QK_n$ with the previous node to extract the key $QK_1$
\begin{align}
\bigoplus_{i=1}^{n-1} x_i \oplus QK_n 
&= \left(QK_1 \oplus QK_2\right) \oplus \left(QK_2 \oplus QK_3\right) \oplus \ldots \nonumber\\
&\quad \oplus \left(QK_{n-1} \oplus QK_n\right) \oplus QK_n \nonumber\\
&= QK_1,
\label{eq:xor}
\end{align}
The adversary can only learn the key if they gain physical access to the relay node during the XOR operation or before the QKD keys are dismissed, as the keys are discarded immediately after use.

The communication schemes rely on all the intermediate nodes to be trusted, and the key becomes totally insecure if any of the nodes become compromised. This problem is resolved by multi-path quantum networking.

\subsection{multi-path Quantum Networking} \label{sec:KDF}
The trusted node network ensures security when all the nodes in a network are trusted. However, for a network comprising multiple paths between two users, secure communication is ensured even if one path is secure. 

\begin{figure}[tbp]
    \centering       \includegraphics[scale=0.3]{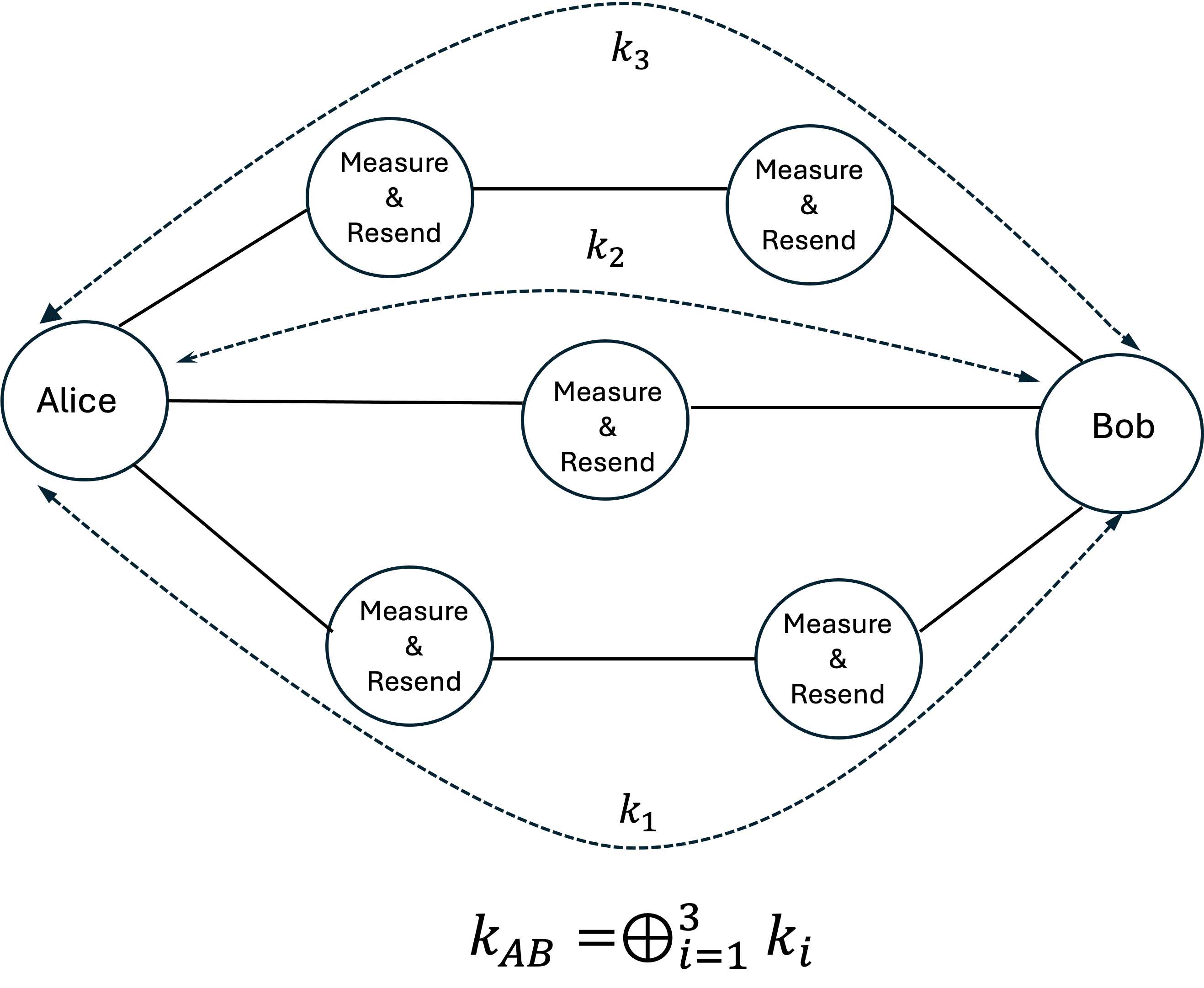}      
    \caption{A trusted node multi-path quantum relay network where keys $k_i$ through each $i$th path are independently distributed. The final key, $k_{AB}=k_1\oplus k_2 \oplus \dots \oplus k_n$ is formed by hashing together $n$ independent keys.} 
    \label{fig:multi_path_QKD}
\end{figure}

In a multi-path QKD network multiple independent paths, all of which rely on trusted nodes for forwarding, are utilized and the independently distributed secret keys are hashed together using a cryptographic hash function to form the final key as shown in Fig.~\ref{fig:multi_path_QKD}. For a given desired key $k$ of length $|k|$ bits, a set of $n$ independent keys $\{k_i\}$, all of the length $|k|$, can be combined into a single key via a bit-wise XOR (addition modulo 2) operation~\cite{salvail2010security},
\begin{align} 
    k = \bigoplus_{i=1}^n k_i,
    \label{eq:KDF1}
\end{align}
where we assume all keys are random bit-strings and therefore of maximum entropy.

The key derivation function (KDF) in Eq.~(\ref{eq:KDF1}), has the property that no information about $k$ can be known unless \emph{all} of the constituent $\{k_i\}$ are known. To see this, let us assume that all but one of the $\{k_i\}$ are known, say $k_1$. Then we can express Eq.~(\ref{eq:KDF1}) as,
\begin{align} \label{eq:KDF}
    k &= k_1 \bigoplus_{i=2}^n k_i,\nonumber\\
    &= k_1 \oplus k_\mathrm{known},
\end{align}
where the term $k_\mathrm{known}$ is a known bit-string. Hence, $k$ is now a known bit-string XORed with an unknown random bit-string, for which the corresponding $k$ is equally likely to be any of the $2^{|k|}$ possibilities with maximum entropy. Intuitively, $k$ has effectively been one-time-pad encrypted with the unknown $k_1$,
making it information-theoretically impossible to determine without knowledge of $k_1$.

multi-path QKD exploits this property of the XOR KDF such that each $k_i$ is established via an independent route through the network of trusted nodes, immediately leading to the observation that if multiple secret keys $k_i$ shared via independent routes are combined, the final derived key $k$ is secure so long as at least one path was uncompromised.

\section{Multi-path quantum Networking with Graph Theoretic Analysis}\label{Sec:Multipath_sec}
The ability of the multi-path networking to ensure security with at least one path being secure makes it a viable solution for secure communication. We build graph-theoretic approach based on statistical model of multi-path network to analyze multi-path networking and show that security can be enhanced by adding redundant edges in the network.

 In Sec.~\ref{sec:Multipathmodel}, we give a statistical model of a multi-path network, relating the security of network with that of independent paths. In Sec.~\ref{sec:grapghtheoretic} we introduce graph theoretic approach to finding the most secure independent path-sets in a multi-path network.

\subsection{Statistical Multi-path Model}\label{sec:Multipathmodel}

Based on the observations for the security of keys derived from multiple independent keys in Sec.~\ref{sec:KDF}, we are able to construct security criteria for multi-path QKD.

Let us assume that all nodes $i$ in the network independently have probability $P_\mathrm{node}(i)$ of being secure (i.e not compromised). For a single path of length $L$ (measured by the number of nodes between the endpoints), the path is secure so long as \emph{all} trusted nodes within it are secure. Hence the probability of a path comprising nodes $\vec{w}$ being secure is,
\begin{align} \label{eq:path_prob}
    P_{\rm path}(\vec{w}) = \prod_{i=1}^L P_{\rm node}(w_i)
\end{align}
where $\omega_i\in\vec{\omega}$.

Next, consider a set of $N$ paths $p_i\in\vec{p}$, with no common vertex, through a network. From Sec.~\ref{sec:KDF} this set of paths yields a secure key unless \emph{all} paths are compromised, with probability of being secure 
\begin{align}
    P_{\rm net}(\vec{p}) = 1-\prod_{i=1}^N [1-P_{\rm path}(p_i)].
\end{align}

A simplified model is when all nodes have an equal probability $P_{\rm node}(i)=P_{\rm node}$ of being secure, and therefore the multi-path is parameterized only by the set of path lengths $\vec{L}$
\begin{align}\label{eq:P_{net}}
    P_{\rm path}(L) &= P_{\rm node}^{L},\nonumber\\
    \end{align}
    such that
    \begin{align}
    P_{\rm net}(\vec{L}) &= 1 - \prod_{i=1}^{|\vec{L}|} (1-P_{\rm path}(L_i))\nonumber\\
    &= 1 - \prod_{i=1}^{|\vec{L}|} (1-P_{\rm node}^{L_i})
\end{align}
is the probability of all paths being secure.
\subsection{Graph Theoretic Approaches for Multi-path networking}\label{sec:grapghtheoretic}
We now introduce our graph theoretic approach to simulate a network. A graph \(G=(V, E)\) is a mathematical structure composed of a vertex set \(V\) and an edge set E, where each edge \(e\epsilon E\) is a pair of vertices \{u,v\}. The order of \(G\), denoted \(|G|\), is the cardinality of \(|V|\), i.e.,\( |G|=|V| \). 

In Sec. \ref{sec:IndependentPath} we provide an efficient algorithm to find the most secure independent path-sets for multi-path networks. In Sec. \ref{sec: Tsecurity} we give our approach to finding a T-secure path set when T-compromised nodes are present at an unknown location in network.

\subsubsection{Finding the most secure independent path-sets}\label{sec:IndependentPath}

Suppose we wish to find the single most secure path between two endpoints in a network. This can be found directly by considering conventional shortest-path algorithms such as Dijkstra's algorithm~\cite{dijkstra1959note} which has efficient $O(|G|^2)$ scaling. From Eq.~(\ref{eq:path_prob}) we note that the probability of having security is the product of the security of the nodes within it. By transforming these probabilities into the logarithmic form they become additive,
\begin{align}
    \log(P_\mathrm{path}) = \sum_{e\in\mathrm{path}} \log(P_e).
\end{align}
Thus we replace the node probabilities $p$ with $-\log(p)$ to make them positive and additive. Now a shortest-path algorithm based on the node-security-weighted graph will yield the most secure route.

To generalize this to multi-path QKD we can iteratively utilize the shortest-path algorithm, removing previously utilized nodes from the graph after each iteration. This yields an efficient greedy algorithm, \texttt{GreedyPaths()}, shown in Alg.~\ref{alg:greedy}.

\begin{algorithm}[tbp]
 \DontPrintSemicolon
 \SetKwInOut{Input}{input}
 \SetKwInOut{Output}{output}
 \SetKwFunction{Func}{GreedyPaths}
 \SetKwFunction{ShortestPath}{ShortestPath}
 \Input{Graph $G$, source $A$, destination $B$, maximum paths $T$.}
 \Output{Path set.}
 \BlankLine
 \Begin{
 	$\mathtt{paths} = \emptyset$\;
 	\For{$n\leftarrow 1$ \KwTo $T$}{
  	   	$\mathtt{path}$ = \ShortestPath($G$,$A$,$B$)\;
        \If{$\mathtt{path}=\emptyset$}{
 			\texttt{error}(``insufficient paths'')\;
 		}
 		$\mathtt{paths} = \mathtt{paths} \cup \mathtt{path}$\;
 		$G = G / \mathtt{path}$\;
 	}
	\Return(\texttt{paths})
}
\caption{\texttt{GreedyPaths()} algorithm finds the most secure disjoint set of $T$ paths with runtime $O(T\cdot|G|^2)$.} \label{alg:greedy}
\end{algorithm}
\subsubsection{Finding $T$-secure path-sets}\label{sec: Tsecurity}
In \cite{salvail2010security} the notion of $T$-security was introduced, whereby a network is guaranteed to be secure if there are \emph{at most} $T$ compromised nodes. Given that it is not known to us which nodes are compromised (otherwise we'd trivially avoid using them), how do we achieve $T$-security in a given network? An algorithm for achieving this, \texttt{SecurePathSet()}, is shown in Alg.~\ref{alg:sec_path_set}. Conceptually, we consider every possible combination of which $T$ distinct nodes can occur in the graph $G$, and for each combination find a path that avoids the respective compromised nodes. If the algorithm is successful, this set of paths, therefore, contains at least one that avoids every possible combination of compromised nodes. Note however that the algorithm is inefficient in $T$, since the number of combinations scales as,
\begin{align}
    \binom{|G|-2}{T},
\end{align}
where $|G|$ is the number of vertices in the graph, discounted by the two endpoints which are assumed to be secure.
\\

\begin{algorithm}[tbp]
 \DontPrintSemicolon
 \SetKwInOut{Input}{input}
 \SetKwInOut{Output}{output}
 \SetKwFunction{Func}{GreedyPaths}
 \SetKwFunction{ShortestPath}{ShortestPath}
 \Input{Graph $G$, source $A$, destination $B$, maximum compromised nodes $T$.}
 \Output{Path set.}
 \BlankLine
 \Begin{
 	$\mathtt{paths} = \emptyset$\;
 	$\mathtt{compnodes} = \bigcup\limits_{\vec{v}\in G, |\vec{v}|=T}\vec{v}$\;
 	\For{$c\in \mathtt{compnodes}$}{
 	    $G' = G / c$\;
  	   	$\mathtt{path}$ = \ShortestPath($G'$,$A$,$B$)\;
      	$\mathtt{paths} = \mathtt{paths} \cup \mathtt{path}$\;
 	}
	\Return($\mathtt{paths}$)
}
\caption{\texttt{SecurePathSet()} algorithm finds a path-set guaranteed to contain a secure path given at most $T$ compromised nodes with runtime $O(|G|^T)$.} \label{alg:sec_path_set}
\end{algorithm}
 \section{Diverse Multi-path Networks Security Analysis}\label{sec:applytopologies}Quantum networks can be of diverse topologies. We extend our multi-path network security analysis to various network topologies. We apply an iterative method to add optimal edges to various network topologies to enhance security. We analyze the resilience of different networks, with some compromised nodes, against hacking.  
 
 In Sec.~\ref{sec:topologies}, we introduce various network topologies, examining their structures. In Sec.~\ref{sec: Analysis}, we evaluate our model's performance across these different network topologies, analyzing how each configuration impacts the efficiency and effectiveness of the network. In Sec.~\ref{sec:optimaledge}, we apply an iterative method to identify and add optimal edge to each topology which gives maximum gain in security. In Sec.~\ref{sec:Phack} we compute the hacking probability of various networks when some nodes are compromised. 
 \subsection{Network Topologies}\label{sec:topologies}
 Network topologies define the arrangement and interconnection of nodes within a network. Designing reliable and efficient networks that meet certain requirements requires an understanding of these topologies. The topologies we explore include Line, Wheel, Star, Grid, Cycle, Tree and Connected-Tree shown in Table~\ref{fig:topologies}.\\
  \begin{table}[tbp]
    \centering
    \begin{tabular}{|m{2.4cm}|m{2.4cm}|m{2.4cm}|} 
        \hline
        \includegraphics[width=2.4cm]{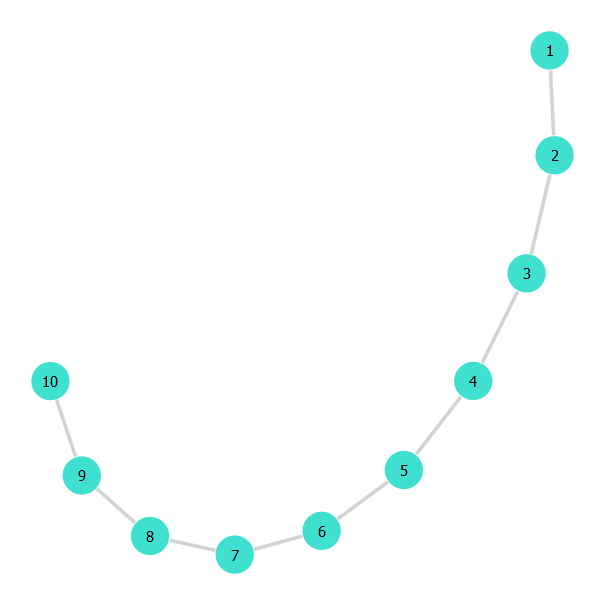}
        \caption*{Linear}&  \includegraphics[width=2.4cm]{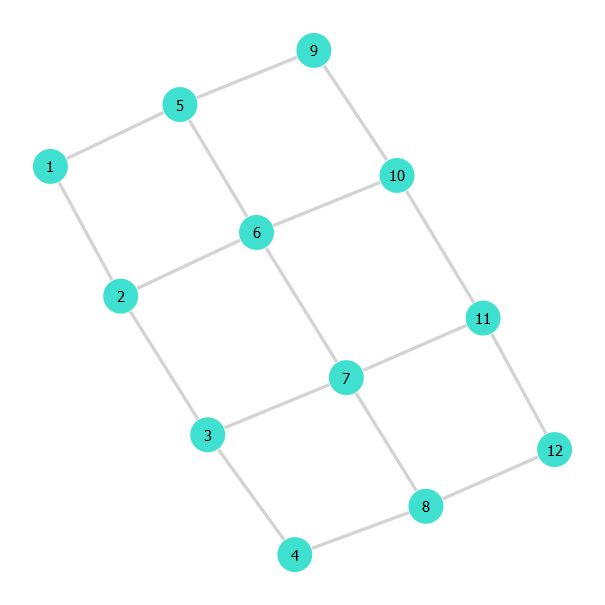}\caption*{Grid} & \includegraphics[width=2.4cm]{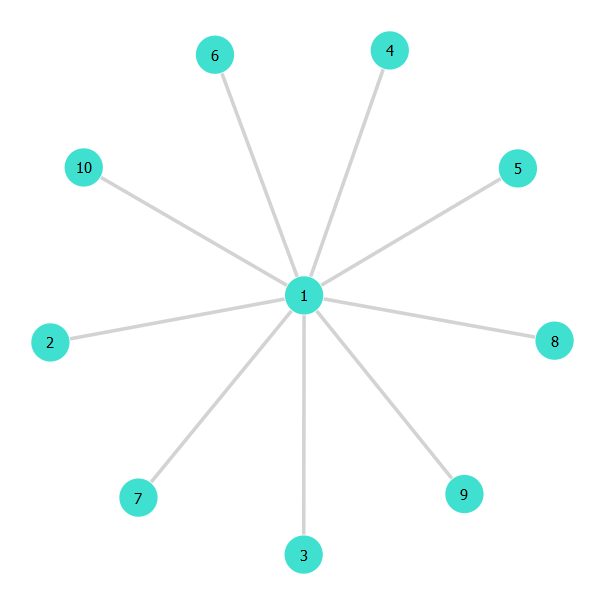}\caption*{Star}\\ 
        \hline
        \includegraphics[width=2.4cm]{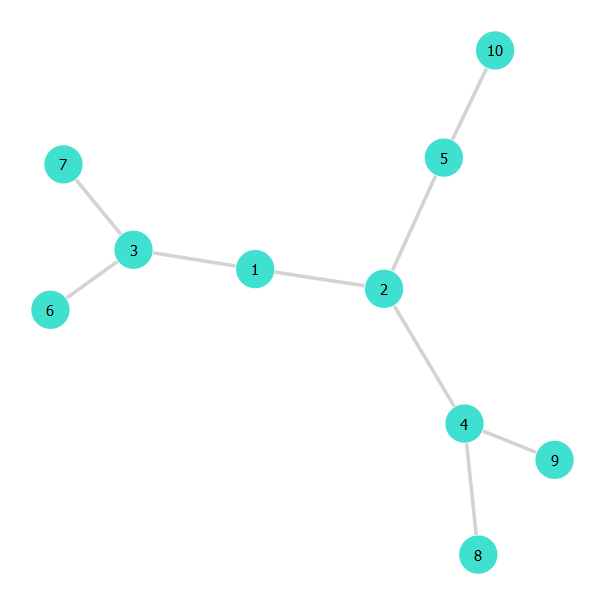}\caption*{Tree} & \includegraphics[width=2.4cm]{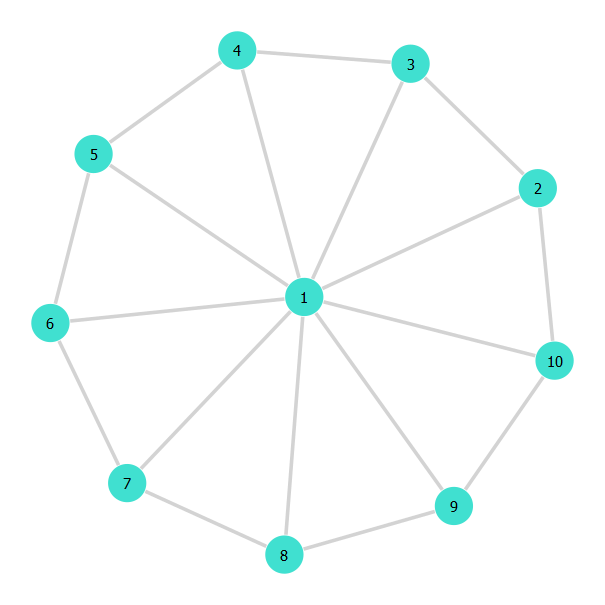}\caption*{Wheel} & \includegraphics[width=2.4cm]{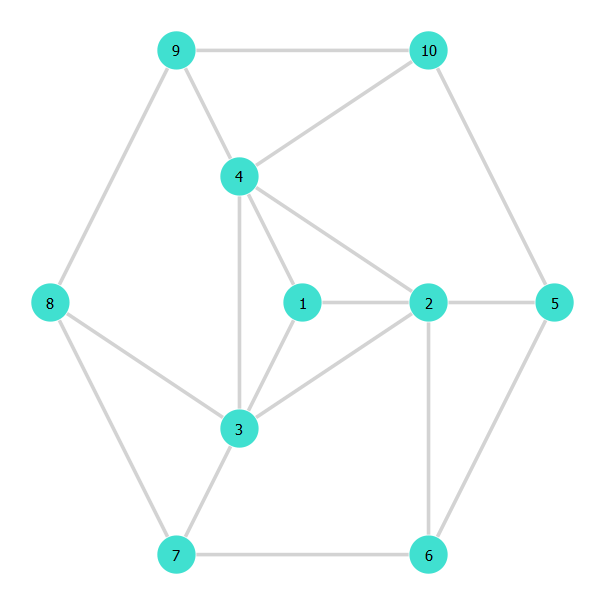}\caption*{Connected-Tree} \\ 
        \hline
         \includegraphics[width=2.4cm]{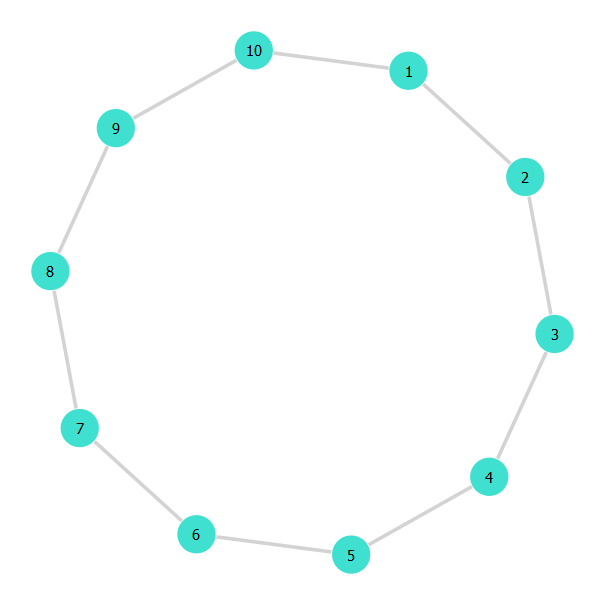}\caption*{Cycle} &  &  \\
         \hline
    \end{tabular}
    \caption{ Various network Topologies like linear, tree, cycle, grid, star, wheel, connected tree. }
    \label{fig:topologies}
\end{table}
  In Line Topology, nodes are connected in a sequential manner where each node (except the endpoints) is connected to two other nodes. A Wheel Topology consists of a central hub node connected to all other peripheral nodes, which also form a closed loop among themselves. Star Topology features a central hub node connected directly to all other nodes, with peripheral nodes not connected to each other. In Grid Topology nodes are arranged in a two-dimensional grid structure where each node is connected to its immediate neighbours horizontally and vertically. In Cycle Topology all nodes are connected in a closed loop, with each node connected to two other nodes forming a ring.
Tree Topology is a hierarchical structure with a single root node connected to one or more subordinate nodes, which in turn may have their own subordinate nodes. Connected Tree Topology is an extension of the tree topology and ensures additional connections to enhance redundancy while maintaining a hierarchical structure.
\subsection{Security Analysis}\label{sec: Analysis}
We evaluate a diverse set of network topologies, each consisting of ten nodes to maintain consistency across comparisons. 
\begin{figure}[tbp]
    \centering
    \includegraphics[scale=0.6]{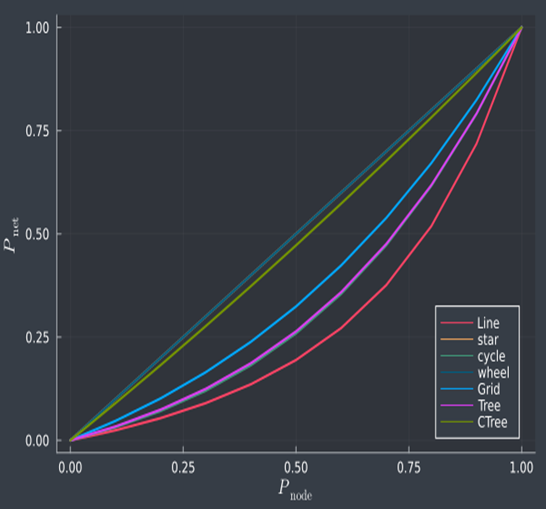}\\
    \includegraphics[scale=0.6]{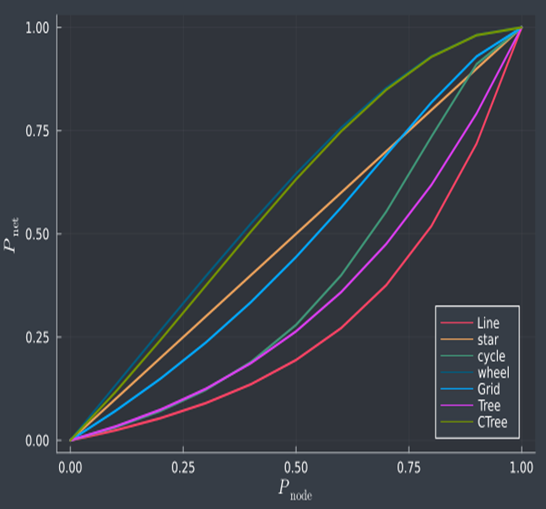}\\
    \includegraphics[scale=0.6]{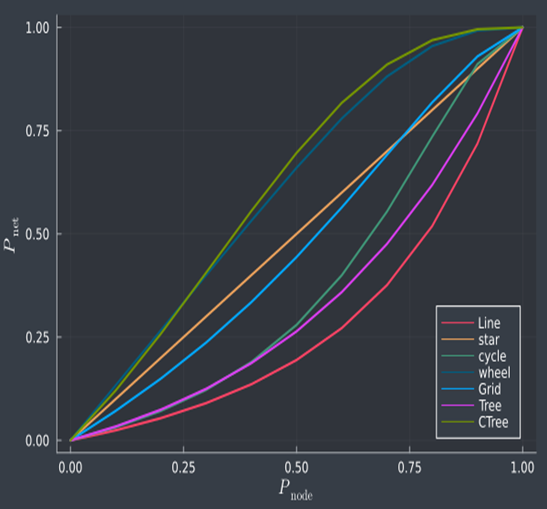}
    \caption{Probability of having security $P_\text{net}$ vs the node security $P_\text{node}$ is obtained from \texttt{GreedyPaths()} for maximum 1 path, 2 paths and 3 paths from top to bottom.}
    \label{fig:Psec_Pnodes}
\end{figure}
We can see in Fig.~\ref{fig:Psec_Pnodes} that different network topologies exhibit unique \(P_{\rm net} \) vs. \(P_{\rm node}\)  relationships, particularly when the path between a source and destination is limited to one. This uniqueness is because every topology has a different number of intermediate nodes between the user pairs. Some topologies, like connected tree and wheel graphs, inherently provide better security, reflected in their initial \(P_{net}\) values for single-path connections. As the number of pathways increases, these topologies experience significant security improvements due to their ability to support multiple connections, which enhances resilience against single points of failure. Conversely, topologies with only a single path between nodes, such as line, star, and tree graphs, show no improvement in security as the path count increases. This is because a single compromised node can jeopardize the entire path. The key finding is that topologies with higher interconnectivity and multiple paths exhibit marked security enhancements as the number of paths increases, underscoring the importance of network topology in determining a network’s security and resilience against attacks or disruptions using a multi-path approach.
 \subsection{Identifying and Implementing the Optimal Edge}\label{sec:optimaledge}
In the analysis of various network topologies, the relationship between network performance (average of \(P_{\rm net}\) for every user pair in the network) and node performance or security of node (\(P_{\rm node}\)) was examined. To enhance the average of \(P_{\rm net}\), an iterative method was employed as follows: Initially, the avg \(P_{\rm net}\) versus \(P_{\rm node}\) relationship was plotted for the given topologies. Subsequently, the complement of each graph, encompassing all potential edges, was generated. Edges were then iteratively selected from this complement graph and added to the original graph. For each edge addition, the impact on the avg \(P_{\rm net}\) versus \(P_{\rm node}\) relationship was assessed. The edge that yielded the most substantial improvement in \(P_{net}\) was identified. This iterative process is governed by Algorithm~\ref{Alg: Optimal_edge}.
\begin{algorithm}[tbp]
 \DontPrintSemicolon
 \SetKwInOut{Input}{input}
 \SetKwInOut{Output}{output}
 \SetKwFunction{Benchmark}{benchmark}
 \SetKwFunction{NetStats}{net\_stats}
 \SetKwFunction{AddEdge}{add\_edge!}
 \SetKwFunction{DeepCopy}{deepcopy}
 \Input{Graph $G$, maximum paths $T$, probability $p$.}
 \Output{Graphs with best average, best lower bound, and best upper bound edges.}
 \BlankLine
 \Begin{
    $gc \gets \text{graph\_complement}(G)$\;
    
    $best\_av \gets 0$; $best\_av\_edge \gets \text{nothing}$\;
    $best\_lower \gets 0$; $best\_lower\_edge \gets \text{nothing}$\;
    $best\_upper \gets 0$; $best\_upper\_edge \gets \text{nothing}$\;
    
    \For{each edge in edges($gc$)}{
        $g\_test \gets \DeepCopy(G)$\;
        \AddEdge{$g\_test$, edge}\;
        $M \gets \Benchmark(G, T)$\;
        $(av, lower, upper) \gets \NetStats(g\_test, M, p)$\;
        
        \If{$av > best\_av$}{
            $best\_av \gets av$\;
            $best\_av\_edge \gets edge$\;
        }
        \If{$lower > best\_lower$}{
            $best\_lower \gets lower$\;
            $best\_lower\_edge \gets edge$\;
        }
        \If{$upper > best\_upper$}{
            $best\_upper \gets upper$\;
            $best\_upper\_edge \gets edge$\;
        }
    }
    
    $g\_best\_av \gets \DeepCopy(G)$\;
    $g\_best\_lower \gets \DeepCopy(G)$\;
    $g\_best\_upper \gets \DeepCopy(G)$\;
    
    \AddEdge{$g\_best\_av$, $best\_av\_edge$}\;
    \AddEdge{$g\_best\_lower$, $best\_lower\_edge$}\;
    \AddEdge{$g\_best\_upper$, $best\_upper\_edge$}\;

    \Return{($g\_best\_av$, $g\_best\_lower$, $g\_best\_upper$)}\;
}
\caption{\texttt{optimal\_edge()} function determines the best edges to add based on average, lower, and upper-performance bounds.}\label{Alg: Optimal_edge}
\end{algorithm}
The impact of each added edge on the overall security posture of the network was evaluated through a comparative analysis of avg \(P_{\rm net}\) versus \(P_{\rm node}\). This method facilitates the identification of critical edges that significantly boost average \(P_{\rm net}\) across various network topologies. 

In Table:~\ref{table:optimaledge}, we present our results for identification and addition of optimal edges.
The table pinpoints individual edges that effectively enhance network security, as measured by network performance factor avg \(P_{\rm net}\). This underscores the importance of strategic edge selection in fortifying network security and offers insights into the network’s vulnerability and robustness.
The identified edges play a crucial role in strengthening the network’s security and resilience.
\begin{table}[tbp]
    \centering
    \begin{tabular}{|m{2.5cm}|m{2.5cm}|m{2.5cm}|} 
        \hline        \includegraphics[width=2.5cm]{Topoligies/Line.png} & \includegraphics[width=2.5cm]{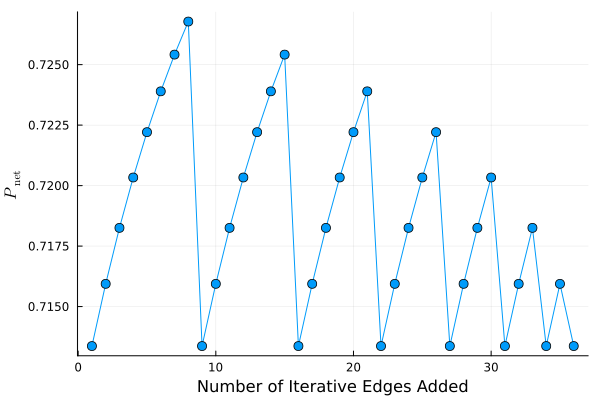}& \includegraphics[width=2.5cm]{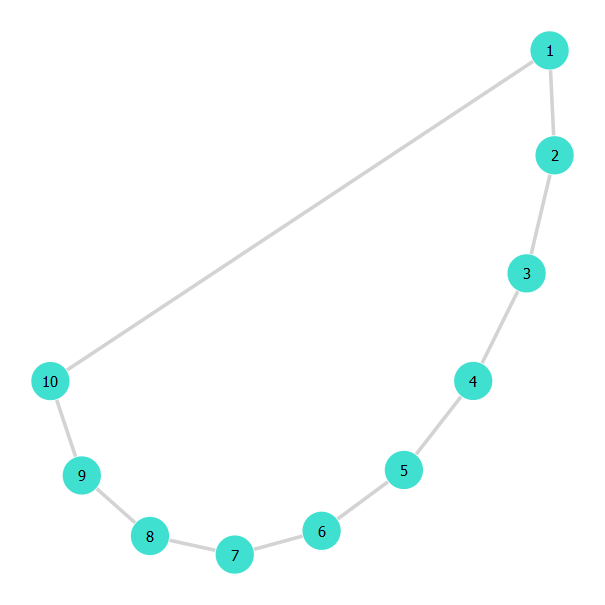}\\ 
        \hline        \includegraphics[width=2.5cm]{Topoligies/TreeGraph.png} & \includegraphics[width=2.5cm]{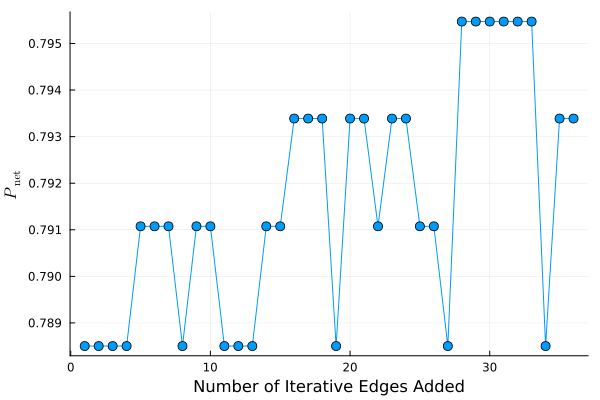} & \includegraphics[width=2.5cm]{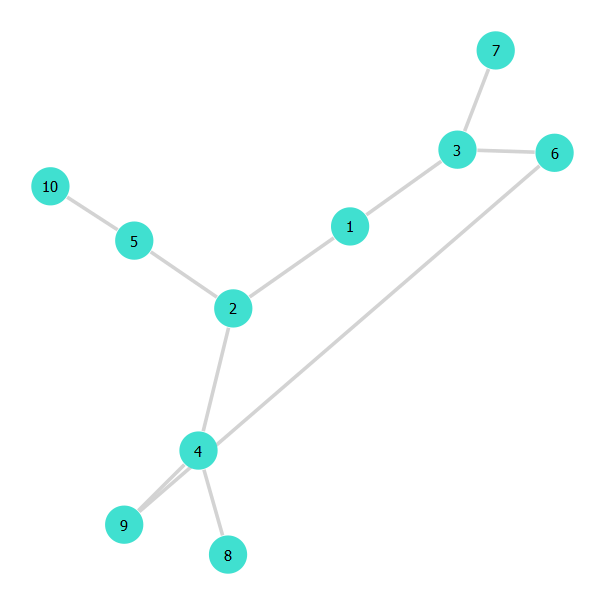} \\ 
        \hline         \includegraphics[width=2.5cm]{Topoligies/Cycle.png} & \includegraphics[width=2.5cm]{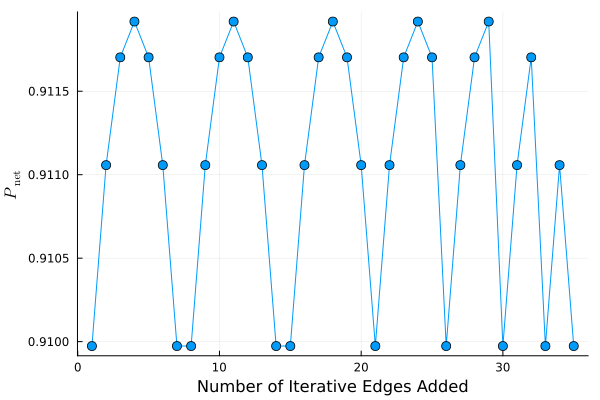} & \includegraphics[width=2.5cm]{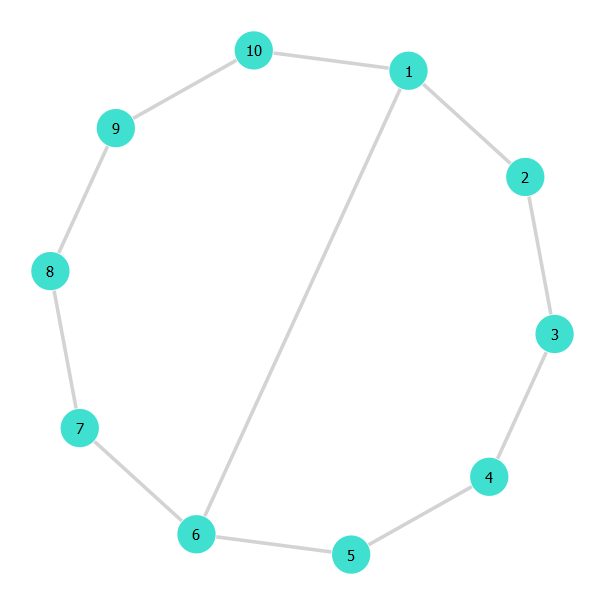}  \\
         \hline        \includegraphics[width=2.5cm]{Topoligies/Grid.png} & \includegraphics[width=2.5cm]{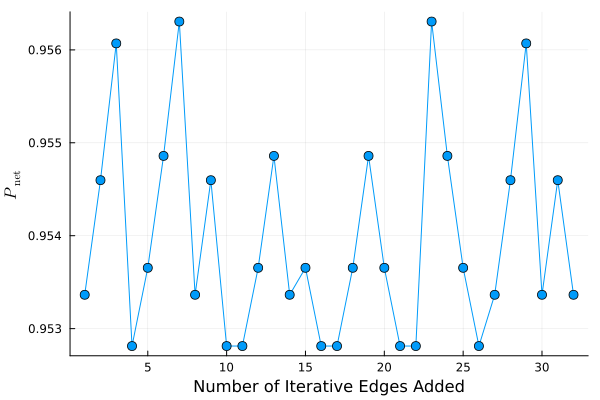} & \includegraphics[width=2.5cm]{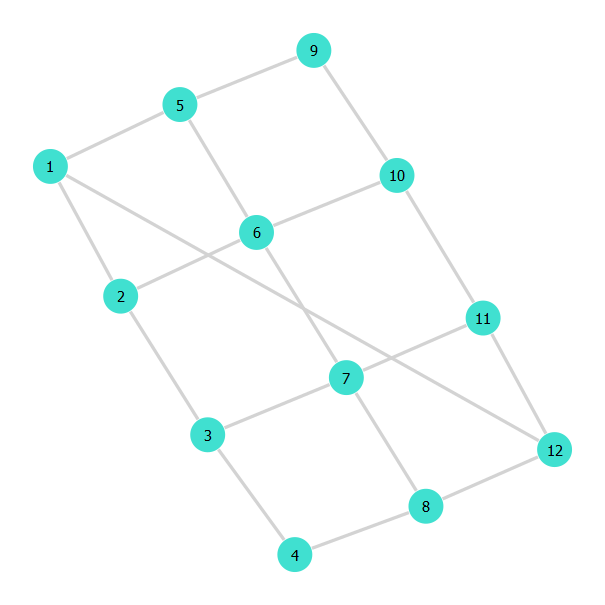}\\
        \hline         \includegraphics[width=2.5cm]{Topoligies/Star.png} & \includegraphics[width=2.5cm]{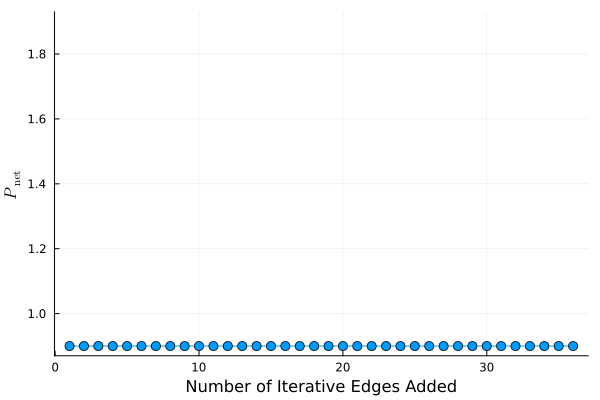} & \includegraphics[width=2.5cm]{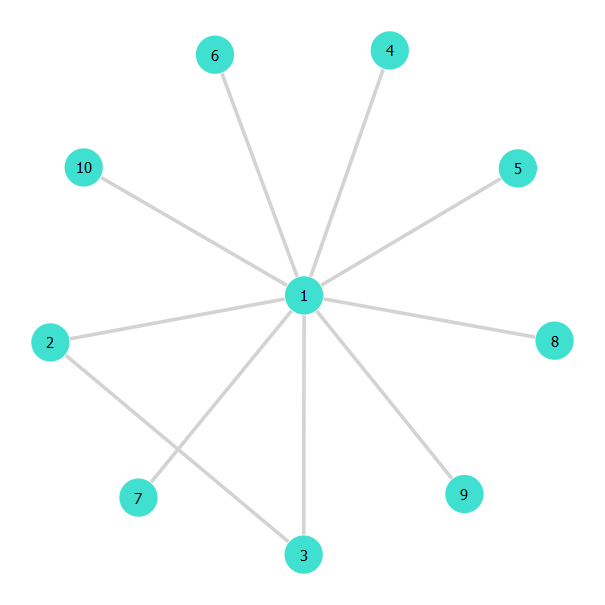} \\ 
         \hline        \includegraphics[width=2.5cm]{Topoligies/wheel.png} & \includegraphics[width=2.5cm]{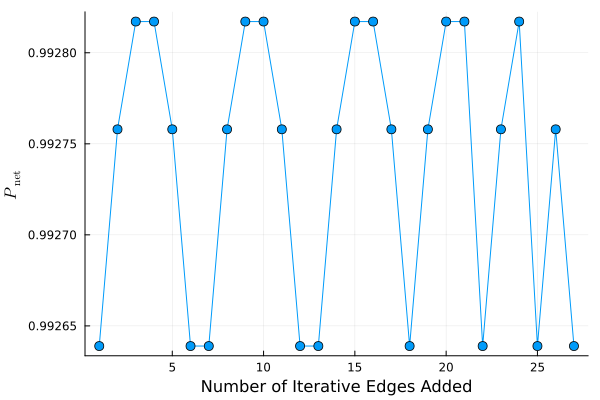} & \includegraphics[width=2.5cm]{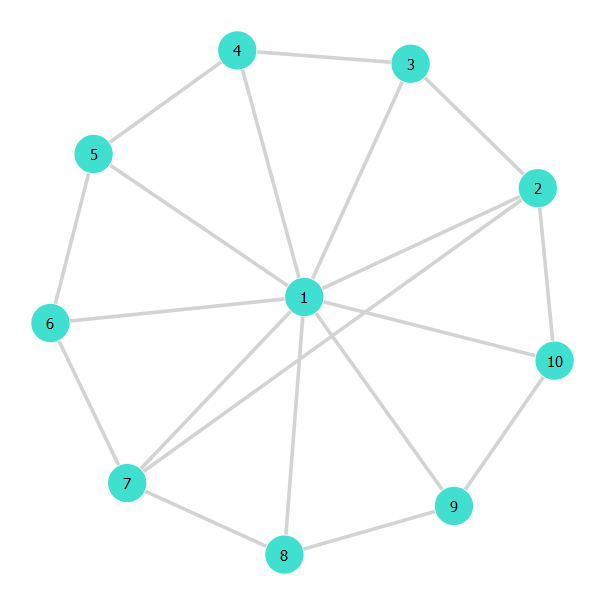}\\ 
        \hline
        \includegraphics[width=2.5cm]{Topoligies/CTreeO.png} & 
         \includegraphics[width=2.5cm]{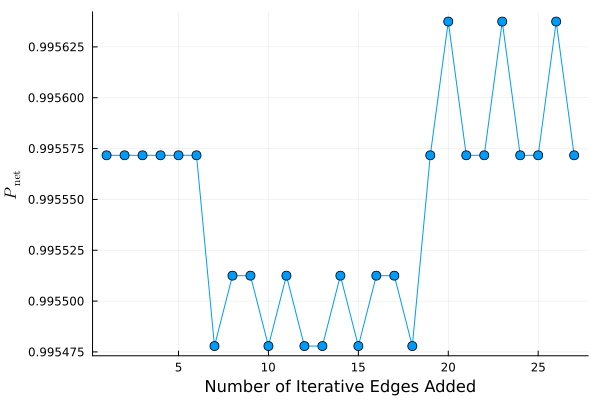} & \includegraphics[width=2.5cm]{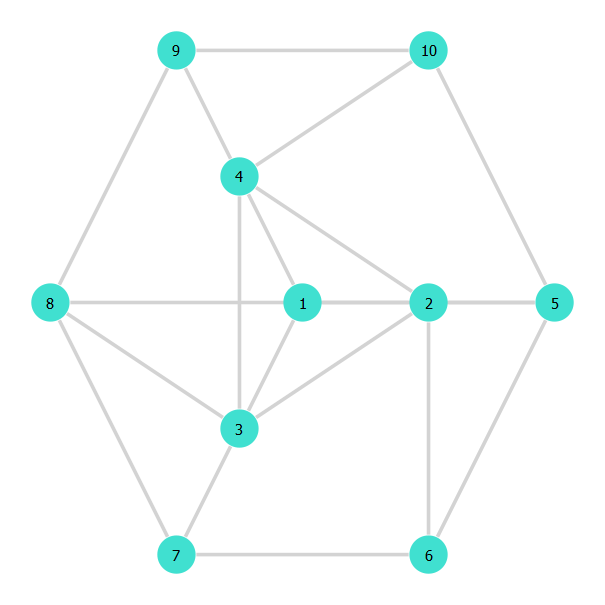} \\
         \hline
    \end{tabular}
    \caption{ Various network Topologies like linear, tree, cycle, grid, star, wheel and connected tree with their corresponding network security and the optimal edge . }
    \label{table:optimaledge}
\end{table}

\subsection{Networks Probability of being hacked \(P_{hack}\)}\label{sec:Phack}
In this study, we analyze the security robustness of various network topologies by examining the probability of a network hack occurring when certain nodes within the network are compromised. The following sections detail the methodology employed to conduct this analysis.
The analytical procedure begins with the identification of all possible secure paths between source and destination for each network topology under consideration. This involves enumerating all unique paths that do not revisit any node and determining their availability based on the network's specific structure. Next, the distribution of compromised nodes \(T_{\rm nodes}\) is systematically varied from 0 up to \(|G| - 2\), excluding the source and destination nodes from being compromised. For each value of \(T_{\rm nodes}\) , all possible combinations of compromised nodes are evaluated.
\begin{align}
P_{hack} =\frac{N_c}{^{|G|-2}C_{T_{\rm nodes}}}
\end{align}

\begin{figure}
    \centering
    \includegraphics[scale=0.5]{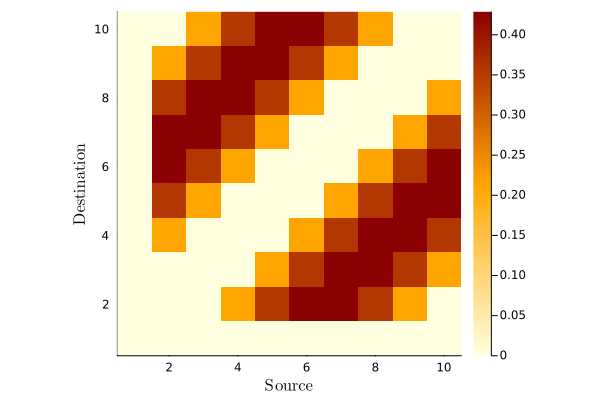}
    \caption{Probability of being hacked as Heat map of wheel graph \(T_{nodes} = 4\). }
    \label{fig:Heat_Map}
\end{figure}

\begin{figure}[tbp]
    \centering
    \includegraphics[scale=0.3]{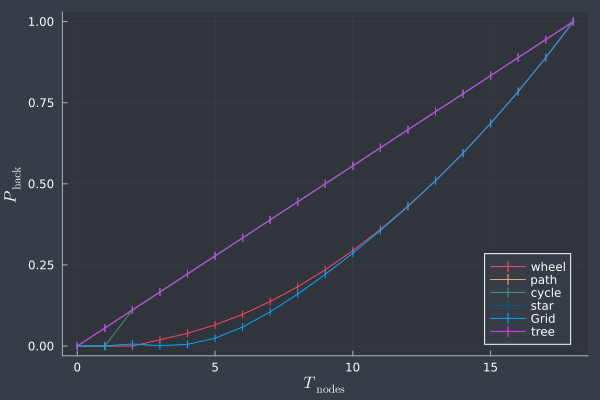}\\
    \includegraphics[scale=0.3]{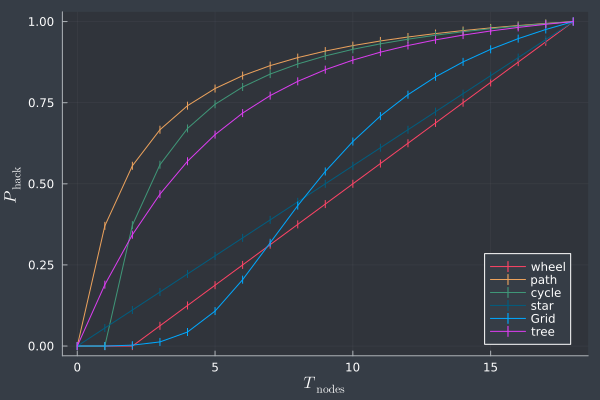}\\
    \includegraphics[scale=0.3]{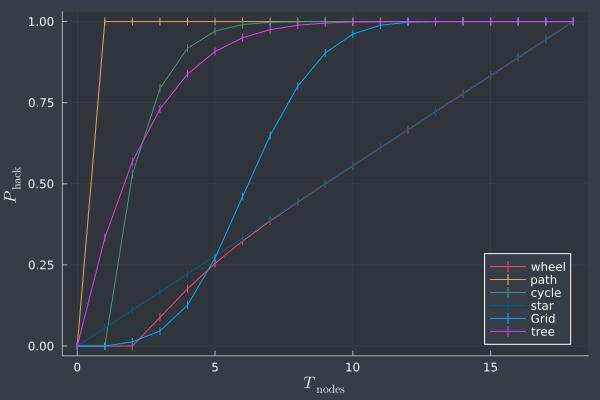}
    \caption{Probability of being hacked $P_\text{hack}$ vs $T_\text{nodes}$ is obtained from \texttt{Hack\_Matrix()} for lower,mean and upper \(P_{hack}\) of network from top to bottom.}
    \label{fig:Psec_Pnodes_Key}
\end{figure}

The secure path evaluation is then performed, where each combination of compromised nodes is assessed to determine if at least one secure path remains between the end user pair. The value of \(N_c > 0\) is the number of combinations for every scenario where no secure path is available due to the compromised nodes. Following this, the probability of a hack (\(P_{\rm hack}\)) is computed for each value of \(T_{\rm nodes}\) across all network topologies, using the obtained \(N_c\) values and the total possible combinations. The Alg.~\ref{Alg:Hack_Matrix} governs this process.
\begin{algorithm}[tbp]
 \DontPrintSemicolon
 \SetKwInOut{Input}{input}
 \SetKwInOut{Output}{output}
 \SetKwFunction{SecurePaths}{secure\_paths}
 \Input{Graph $G$, Integer $T$.}
 \Output{Matrix $H$ representing the probability of compromised communication between nodes.}
 \BlankLine
 \Begin{
    $div \gets 1$\;
    $H \gets \text{zeros}(nv(G), nv(G))$\tcp{Matrix of size $nv(G) \times nv(G)$}\;
        \For{each $src\leftarrow 1$ \KwTo $nv(G)$}{
            \For{each $dst\leftarrow 1$ \KwTo $nv(G)$}{
                \If{$src == dst$}{
                    $H[src, dst] \gets 0$\;
                }
                \Else{
                    $A \gets 0$\;
                    $d \gets \SecurePaths(G, src, dst, T)$\;
                    \For{each $m$ in $d$}{
                        \If{$d[m]$ is empty}{
                            $A \gets A + 1$\;
                        }
                    }
                    $H[src, dst] \gets \frac{A}{\text{length}(d)}$\;
                }
            }
        }
    \Return{$H$}\;
}
\caption{\texttt{Hack\_Matrix()} algorithm computes the probability matrix $H$ representing compromised communication between nodes.}
\label{Alg:Hack_Matrix}
\end{algorithm}
Fig.~\ref{fig:Heat_Map} provides a comprehensive view of network security by visualizing the probability of being hacked when~\(T_{\rm nodes}\) are compromised. Each cell in the heat map represents a source-destination pair, with the color of the cell indicating the likelihood that communication between that pair is compromised. Darker or more intense colors correspond to higher probabilities of being hacked, allowing for a clear and immediate understanding of network vulnerabilities under different compromise scenarios. This visualization helps identify which pairs are more vulnerable to attacks and how the overall network security is affected by compromised nodes.\\
Finally, in Fig.~\ref{fig:Psec_Pnodes_Key} the relationship between \(T_{\rm nodes}\) and \(P_{\rm hack}\) is analyzed to identify trends and critical thresholds that indicate network vulnerability. The results are then visualized using appropriate plots to illustrate how the increase in the number of compromised nodes affects the security of each network topology, providing insights into the resilience of different network structures under varying levels of node compromise.
\section{Key Rate Analysis with multi-path and Trusted Nodes}~\label{Sec:Key_rate}
In a quantum key distribution (QKD) network employing trusted nodes, these nodes are essential for extending the communication distance and maintaining the integrity of the key. While these trusted nodes ensure that the key remains unaltered (i.e., they do not tamper with the key), there is a concern regarding the privacy of the key. Specifically, trusted nodes may inadvertently or deliberately publish parts of the key, leading to potential information leakage. Although the authentication of the key is preserved, the privacy of the key can be compromised in such scenarios.

To address this issue, we propose a multi-path QKD model with trusted nodes, where the final key is obtained by XORing the keys from multiple independent paths between the communicating parties, Alice and Bob. This multi-path approach enhances the overall privacy of the key, as it reduces the impact of any information leakage from a single node. Even if a trusted node in one path leaks some information, the security of the final key is significantly strengthened because the final key is a combination of keys from all paths.

In a traditional QKD system, the key rate is determined by the weakest link in the communication path. For a single path, the key rate \( R_{\text{single}} \) is given by:
\[
R_{\text{single}} = \min \{ R_1, R_2, \dots, R_N \}
\]
where \( R_1, R_2, \dots, R_N \) are the key rates of individual segments in the path, considering factors like channel loss, error rates, and detector efficiency.

In our proposed model, Alice and Bob establish multiple independent paths for QKD, and the final key is obtained by XORing the keys from these paths. If at least one path is secure then net security is
\begin{align}
    P_{\rm net}(\vec{L}) &= 1 
\end{align}.
The upper bound on the key rate for this multi-path system is determined by the minimum key rate among all $n$ paths
\begin{align}
R_{\text{upper}} = \min \{ R_{\text{path1}}, R_{\text{path2}}, \cdots R_{{\rm path}n} \}
\end{align}
where \( R_{{\rm path}i}\) are the key rates for each path.

The overall performance of the network, considering potential information leakage at trusted nodes, is encapsulated in the network performance factor \( P_{\text{net}} \). This factor accounts for path reliability and privacy amplification. It can be expressed as Eq.~(\ref{eq:P_{net}})

\begin{figure}
    \centering
    \includegraphics[width=0.5\linewidth]{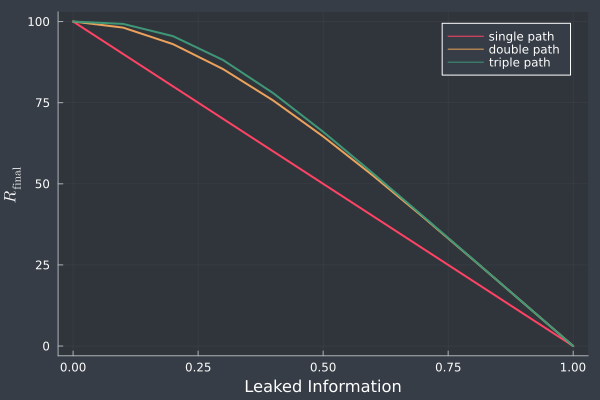}
    \caption{Key rate versus information leakage in a wheel graph.}
    \label{fig: R_Final}
\end{figure}

The final key rate \( R_{\text{final}} \) for the multi-path QKD system is obtained by multiplying the network performance factor \( P_{\rm net} \) with the minimum key rate among the paths. In Fig.~\ref{fig: R_Final} \(R_{\rm final}\) is plotted against information leakage (\(1-P_{\rm node}\)) for single, double, and triple paths where
\[
R_{\text{final}} = P_{\text{net}} \cdot \min \{ R_{\text{path1}}, R_{\text{path2}}, R_{\text{path3}} \}
\]
We see that as the number of paths increases, the key rate significantly increases. This validates the advantage of multi-path QKD with trusted nodes.
\section{Conclusions}~\label{Sec: Conclusion}
Recent experimental advances have demonstrated critical components of a quantum internet, such as quantum repeaters that extend the range of entanglement distribution. These efforts are crucial steps toward creating a global quantum internet, which could revolutionize information processing, computation, and communication in the 21st century.

In this comparative study, we addressed one of the key challenges in realizing the quantum internet: developing effective routing strategies in multi-path networks. Long-distance quantum communication typically relies on quantum repeaters, which are difficult to implement due to the need for quantum memory to store qubits while preserving their entanglement. Advances in quantum memory are essential for practical quantum repeaters.

An alternative approach to long-distance communication involves using trusted nodes instead of quantum repeaters. While trusted nodes do not require quantum memory, they introduce security risks, as compromising even one trusted node could jeopardize the entire communication. To mitigate this risk, we proposed a multi-path Quantum network strategy, which significantly enhances the security of various network topologies.

Our analysis revealed that certain networks have more robust security which further increases when the paths utilized increases. In the case of compromised nodes at unknown locations in the network, our proposed SecurePathSet() algorithm finds the secure paths but it grows exponentially with an increase in compromised nodes. When the degrees of the source and destination nodes exceed the number of compromised nodes, communication is 100\% secure. However, when the number of compromised equals or exceeds the degrees of the source and destination nodes, the probability of a hack depends on different combinations of compromised nodes. Additionally, we identified and implemented the optimal edge for existing topologies to further enhance security.

The proposed multi-path QKD model with trusted nodes
significantly enhances the privacy of the final key by
mitigating the impact of information leakage. By XORing
the keys from multiple paths and incorporating a network
performance factor, the model ensures that the final key
rate remains secure and reliable, even in the presence
of information leakage at trusted nodes. The multi-path
approach not only preserves the authentication of the key
but also provides a more robust defense against potential
privacy breaches, making it a valuable enhancement for
QKD networks.

These findings highlight the importance of network topology in ensuring secure quantum communication and contribute to the ongoing efforts to develop a robust and resilient Quantum Internet.

\section*{Acknowledgments}
Authors thank Peter P. Rohde for many insightful discussions.
\bibliographystyle{spphys}
\bibliography{AAMy}

@article{bennett2014,
  author       = {Charles H. Bennett and Gilles Brassard},
  title        = {Quantum Cryptography: Public Key Distribution and Coin Tossing},
  journal      = {Theoretical Computer Science},
  volume       = {560},
  pages        = {7--11},
  year         = {2014},
  note         = {Theoretical Aspects of Quantum Cryptography -- Celebrating 30 years of BB84},
}

@article{ekert1991quantum,
  title={Quantum cryptography based on Bell’s theorem},
  author={Ekert, Artur K},
  journal={Physical review letters},
  volume={67},
  number={6},
  pages={661},
  year={1991},
  publisher={APS}
}

@article{gisin2002quantum,
  title={Quantum cryptography},
  author={Gisin, Nicolas and Ribordy, Gr{\'e}goire and Tittel, Wolfgang and Zbinden, Hugo},
  journal={Reviews of modern physics},
  volume={74},
  number={1},
  pages={145},
  year={2002},
  publisher={APS}
}

@article{briegel1998quantum,
  title={Quantum repeaters: the role of imperfect local operations in quantum communication},
  author={Briegel, H-J and D{\"u}r, Wolfgang and Cirac, Juan I and Zoller, Peter},
  journal={Physical Review Letters},
  volume={81},
  number={26},
  pages={5932},
  year={1998},
  publisher={APS}
}

@article{kimble2008quantum,
  title={The quantum internet},
  author={Kimble, H Jeff},
  journal={Nature},
  volume={453},
  number={7198},
  pages={1023--1030},
  year={2008},
  publisher={Nature Publishing Group}
}

@article{dong2019wide,
  title={Wide area key distribution network based on a quantum key distribution system},
  author={Dong, Hua and Song, Yaqi and Yang, Li},
  journal={Applied Sciences},
  volume={9},
  number={6},
  pages={1073},
  year={2019},
  publisher={MDPI}
}

@article{salvail2010security,
  title={Security of trusted repeater quantum key distribution networks},
  author={Salvail, Louis and Peev, Momtchil and Diamanti, Eleni and All{\'e}aume, Romain and L{\"u}tkenhaus, Norbert and L{\"a}nger, Thomas},
  journal={Journal of Computer Security},
  volume={18},
  number={1},
  pages={61--87},
  year={2010},
  publisher={IOS Press}
}

@article{dijkstra1959note,
  author       = {Dijkstra, Edsger W.},
  title        = {A Note on Two Problems in Connexion with Graphs},
  journal      = {Numerische Mathematik},
  volume       = {1},
  number       = {1},
  pages        = {269--271},
  year         = {1959},
  publisher    = {Springer},
  doi          = {10.1007/BF01386390}
}

@inproceedings{mondal2024multipath,
  title={Multipath entanglement purification strategies for quantum networks},
  author={Mondal, Md Sohel and Santra, Siddhartha},
  booktitle={2024 International Conference on Quantum Communications, Networking, and Computing (QCNC)},
  pages={312--319},
  year={2024},
  organization={IEEE}
}

@article{bennett1993teleporting,
  title={Teleporting an unknown quantum state via dual classical and Einstein-Podolsky-Rosen channels},
  author={Bennett, Charles H and Brassard, Gilles and Cr{\'e}peau, Claude and Jozsa, Richard and Peres, Asher and Wootters, William K},
  journal={Physical review letters},
  volume={70},
  number={13},
  pages={1895},
  year={1993},
  publisher={APS}
}

@article{leone2021qunet,
  title={QuNet: Cost vector analysis \& multi-path entanglement routing in quantum networks},
  author={Leone, Hudson and Miller, Nathaniel R and Singh, Deepesh and Langford, Nathan K and Rohde, Peter P},
  journal={arXiv preprint arXiv:2105.00418},
  year={2021}
}

@article{wolters2017simple,
  title={Simple atomic quantum memory suitable for semiconductor quantum dot single photons},
  author={Wolters, Janik and Buser, Gianni and Horsley, Andrew and B{\'e}guin, Lucas and J{\"o}ckel, Andreas and Jahn, Jan-Philipp and Warburton, Richard J and Treutlein, Philipp},
  journal={Physical review letters},
  volume={119},
  number={6},
  pages={060502},
  year={2017},
  publisher={APS}
}

@article{daud2024scaling,
  title={Scaling network topologies for multi-user entanglement distribution},
  author={Daud, Muhammad and Khalique, Aeysha},
  journal={Quantum Information Processing},
  volume={23},
  number={6},
  pages={216},
  year={2024},
  publisher={Springer}
}

@article{elliott2002building,
  title={Building the quantum network},
  author={Elliott, Chip},
  journal={New Journal of Physics},
  volume={4},
  number={1},
  pages={46},
  year={2002},
  publisher={IOP Publishing}
}

@phdthesis{abruzzo2014long,
  title={Long distance quantum key distribution with quantum repeaters},
  author={Abruzzo, Silvestre},
  year={2014},
  school={D{\"u}sseldorf, Heinrich-Heine-Universit{\"a}t, Diss., 2014}
}

@article{ma2020optical,
  title={Optical quantum memory and its applications in quantum communication systems},
  author={Ma, Lijun and Slattery, Oliver and Tang, Xiao},
  journal={Journal of Research of the National Institute of Standards and Technology},
  volume={125},
  year={2020},
  publisher={National Institute of Standards and Technology}
}

@article{elliott2004quantum,
  title={Quantum cryptography},
  author={Elliott, Chip},
  journal={IEEE security \& privacy},
  volume={2},
  number={4},
  pages={57--61},
  year={2004},
  publisher={IEEE}
}

@inproceedings{schartner2009overcome,
  title={How to overcome the'Trusted Node Model'in Quantum Cryptography},
  author={Schartner, Peter and Rass, Stefan},
  booktitle={2009 International conference on computational science and engineering},
  volume={3},
  pages={259--262},
  year={2009},
  organization={IEEE}
}

@inproceedings{caleffi2018quantum,
  title={Quantum internet: From communication to distributed computing!},
  author={Caleffi, Marcello and Cacciapuoti, Angela Sara and Bianchi, Giuseppe},
  booktitle={Proceedings of the 5th ACM international conference on nanoscale computing and communication},
  pages={1--4},
  year={2018}
}

@article{azuma2023quantum,
  title={Quantum repeaters: From quantum networks to the quantum internet},
  author={Azuma, Koji and Economou, Sophia E and Elkouss, David and Hilaire, Paul and Jiang, Liang and Lo, Hoi-Kwong and Tzitrin, Ilan},
  journal={Reviews of Modern Physics},
  volume={95},
  number={4},
  pages={045006},
  year={2023},
  publisher={APS}
}

@article{bechmann505089quantum,
  title={Quantum key distribution with trusted quantum relay.(2018)},
  author={Bechmann-Pasquinucci, H and Pasquinucci, A},
  journal={arXiv preprint quant-ph/0505089}
}

@article{StepniakJISA2023,
title = {Analysis of multiple overlapping paths algorithms for secure key exchange in large-scale quantum networks},
journal = {Journal of Information Security and Applications},
volume = {78},
pages = {103581},
year = {2023},
issn = {2214-2126},
doi = {https://doi.org/10.1016/j.jisa.2023.103581},
url = {https://www.sciencedirect.com/science/article/pii/S2214212623001655},
author = {Mateusz Stępniak and Jakub Mielczarek}
}

@article{Solomons2022PRXQuantum,
  title = {Scalable Authentication and Optimal Flooding in a Quantum Network},
  author = {Solomons, Naomi R. and Fletcher, Alasdair I. and Aktas, Djeylan and Venkatachalam, Natarajan and Wengerowsky, S\"oren and Lon{\v c}ari{'c}, Martin and Neumann, Sebastian P. and Liu, Bo and Samec, {\v Z}eljko and Stip\ifmmode \check{c}\else \v{c}\fi{}evi\ifmmode \acute{c}\else \'{c}\fi{}, Mario and Ursin, Rupert and Pirandola, Stefano and Rarity, John G. and Joshi, Siddarth Koduru},
  journal = {PRX Quantum},
  volume = {3},
  issue = {2},
  pages = {020311},
  numpages = {18},
  year = {2022},
  month = {Apr},
  publisher = {American Physical Society},
  doi = {10.1103/PRXQuantum.3.020311},
  url = {https://link.aps.org/doi/10.1103/PRXQuantum.3.020311}
}

@article{Kiktenko2024Entropy,
  title = {Routing Algorithm Within the Multiple Non-Overlapping Paths’ Approach for Quantum Key Distribution Networks},
  author = {Kiktenko, E. O. and Tayduganov, A.
  and Fedorov, A. K.},
  journal = {Entropy},
  volume = {26},
  issue = {12},
  pages = {1102},
  year = {2024},
  doi = {10.1103/PRXQuantum.3.020311},
  url = {https://doi.org/10.3390/e26121102}
}
\end{document}